\documentclass{emulateapj}

\newcommand{\iso}{{\em ISO}}

\newcommand{\mum}{\ifmmode{\rm \mu m}\else{$\mu$m}\fi}
\newcommand{\er}{\ifmmode{\pm}\else{$\pm$}\fi}

\submitted{Submitted to the Astrophysical Journal, 7 Feb. 2005; revised 6 May, 2005; accepted 23 June, 2005}
\journalinfo{ }

\shorttitle{PAHs in Herbig AeBe Stars}
\shortauthors{Sloan et al.}

\begin{document}

\title{Mid-infrared spectra of PAH emission in Herbig AeBe stars}

\author{
G.~C.~Sloan\altaffilmark{1}, 
L.~D.~Keller\altaffilmark{2},
W.~J.~Forrest\altaffilmark{3},
E.~Leibensperger\altaffilmark{2},
B.~Sargent\altaffilmark{3}, 
A.~Li\altaffilmark{4},
J.~Najita\altaffilmark{5},
D.~M.~Watson\altaffilmark{3},
B.~R.~Brandl\altaffilmark{6},
C.~H.~Chen\altaffilmark{5},
J.~D.~Green\altaffilmark{3},
F. Markwick-Kemper\altaffilmark{7},
T.~L.~Herter\altaffilmark{1},
P.~D'Alessio\altaffilmark{8}, 
P.~W.~Morris\altaffilmark{9}, 
D.~J.~Barry\altaffilmark{1}, 
P.~Hall\altaffilmark{1}, 
P.~C.~Myers\altaffilmark{10}, \&
J.~R.~Houck\altaffilmark{1} }

\altaffiltext{1}{Cornell University, Astronomy Department,
  Ithaca, NY 14853-6801, sloan@isc.astro.cornell.edu}
\altaffiltext{2}{Department of Physics, Ithaca College, Ithaca, NY 14850}
\altaffiltext{3}{Department of Physics and Astronomy, University of 
  Rochester, Rochester, NY 14627-0171}
\altaffiltext{4}{Department of Physics \& Astronomy, University of 
  Missouri-Columbia, Columbia, MO 65211}
\altaffiltext{5}{National Optical Astronomy Observatory, 950 North 
  Cherry Avenue, Tucson, AZ 85719}
\altaffiltext{6}{Sterrewacht Leiden, P.O. Box 9513, 2300 RA Leiden, 
  The Netherlands}
\altaffiltext{7}{Astronomy Department, University of Virginia, P.~O. Box
  3818, Charlottesville, VA 22903}
\altaffiltext{8}{Centro de Radioastronomia y Astrofisica, UNAM, 
  Apartado Postal 3-72 (Xangari), 58089 Morelia, Michoacan, Mexico}
\altaffiltext{9}{NASA {\it Herschel} Science Center, IPAC/Caltech, 
  MS 100-22, Pasadena, CA  91125}
\altaffiltext{10}{Harvard-Smithsonian Center for Astrophysics, 
  60 Garden Street, Cambridge, MA 02138}

\begin{abstract}
We present spectra of four Herbig AeBe stars obtained with
the Infrared Spectrograph (IRS)\footnote{The IRS was a
collaborative venture between Cornell University and Ball
Aerospace Corporation funded by NASA through the Jet
Propulsion Laboratory and the Ames Research Center.} on the
{\it Spitzer Space Telescope}.  All four of the sources show
strong emission from polycyclic aromatic hydrocarbons (PAHs),
with the 6.2~\mum\ emission feature shifted to 6.3~\mum\ and
the strongest C$-$C skeletal-mode feature occuring at 
7.9~\mum\ instead of at 7.7~\mum\ as is often seen.  
Remarkably, none of the four stars have silicate emission. 
The strength of the 7.9~\mum\ feature varies with respect to 
the 11.3~\mum\ feature among the sources, indicating that we 
have observed PAHs with a range of ionization fractions.  The
ionization fraction is higher for systems with hotter and brighter
central stars.  Two sources, HD 34282 and HD 169142, show 
emission features from aliphatic hydrocarbons at 6.85 and 
7.25~\mum.  The spectrum of HD 141569 shows a previously 
undetected emission feature at 12.4~\mum\ which may be related 
to the 12.7~\mum\ PAH feature.  The spectrum of HD 135344, the
coolest star in our sample, shows an unusual profile in the
7--9~\mum\ region, with the peak emission to the red of
8.0~\mum\ and no 8.6~\mum\ PAH feature.
\end{abstract}

\keywords{stars: chemically peculiar --- infrared: stars}

\section{Introduction}

Herbig AeBe stars (HAeBe) are intermediate mass (2--8 
M$_{\sun}$) analogs to the roughly solar-mass T Tauri stars 
\citep{her60,str72}.  
The number of HAeBe stars in the Galaxy depends on the selection
criteria, and as a result different studies have listed as few
as $\sim$50 \citep[Herbig's original list included 26 stars, but 
see][]{mal98} or as many as several hundred \citep{the94}.
One of the more conservative definitions assigns the HAeBe name 
to stars of spectral type A and B (and early F) that have 
broadened atomic emission lines in their optical spectra (hence 
the ``e'' in ``AeBe'') and have infrared emission in excess of a 
purely photospheric spectral energy distribution.  Using these 
criteria, \cite{mal98} identified 45 HAeBe stars and published 
spectral energy distributions (SEDs) from UV to mm wavelengths,
proposing an evolutionary scenario based on mid-infrared colors 
and SED structure.  

\cite{mee01} observed the 14 spatially unresolved sources (at
all wavelengths) from the list by \cite{mal98} with the 
Short-Wavelength Spectrometer (SWS) aboard the {\it Infrared 
Space Observatory (ISO)}.  
They report the detection of polycyclic aromatic hydrocarbons 
(PAHs) in seven of the 14 sources, proposing that the emitting 
PAHs are located in flared disks where the shadow of a thicker 
inner disk cannot attenuate the UV radiation.  \cite{aa04} 
recently presented spectra from the SWS for the entire list of 
sources of \cite{mal98}, and they report the detection of PAH 
emission in many of them, mentioning the possibility that 
differences in the ionization fraction of the PAHs could help 
explain the structure in the 2--20~\mum\ spectra.  They also 
noted that emission from crystalline silicates may confuse the 
issue.  It is also clear from the SWS spectrum of 51 Oph 
\citep[HD 156843;][]{vda01} that emission from gas near the 
stars may play a significant role in shaping the mid-infrared 
spectra of these stars and complicating detailed studies of 
PAH emission in that spectral range. 

\cite{hab04} have analyzed ground-based and ISO observations of 
some 30 HAeBe stars and applied a model in which the material 
in the disk is in vertical hydrostatic equilibrium, is heated 
by the central star, and contains large dust grains in thermal 
equilibrium with the radiation field as well as small grains 
and PAHs that are transiently heated.  They suggest that the 
PAH emission is primarily from the outer disks (R$\sim$100 AU) 
and that strong PAH emission is predominantly from sources 
that have flared disks \citep[as earlier proposed 
by][]{mee01}.  

\cite{ll03} modeled ground-based spectra of HD 141569 in the 
8--13 and 17--25~\mum\ regions using a porous, cometary-type 
dust model consisting of coagulated, but otherwise unaltered 
interstellar grains along with fully ionized PAHs.  They 
predict the shape of the 5--25~\mum\ spectrum, noting that 
there should be little emission from crystalline silicates 
(no more than 10\% by mass fraction).  Since PAHs are a 
significant component of the interstellar medium (ISM), important 
to the energy balance \citep[e.g.][]{atb89}, and are 
apparently necessary to accurately model the emission in HAeBe 
stars, it is time to closely investigate not just their presence, 
but the specific spectral properties of the emission features 
from PAHs around HAeBe stars.

\begin{deluxetable*}{llcclr}
\tablecolumns{6}
\tablewidth{0pt}
\tablenum{1}
\tablecaption{Observing information}
\tablehead{ 
  \colhead{Target} & \colhead{Spectral Type} & 
  \colhead {$U(r)$ ($\times$ $10^5$)\tablenotemark{a}} & 
  \colhead{IRS Campaign} & \colhead{Observation Date (UT)} & \colhead{AOR Key}
}
\startdata
HD  34282 & A0e                    & 2.3  & 13 & 2004 Oct. 10 & 3577856 \\
HD 141569 & A0 Ve\tablenotemark{b} & 2.3  &  4 & 2004 Mar.  3 & 3560960 \\ 
HD 169142 & A5 Ve\tablenotemark{b} & 0.71 &  5 & 2004 Mar. 26 & 3587584 \\ 
HD 135344 & F4 Ve\tablenotemark{b} & 0.16 & 11 & 2004 Aug.  8 & 3580672 \\ 
\enddata
\tablenotetext{a}{$U(r)$ is the radiation field due to the star in 
units of the interstellar radiation field in the Solar neighborhood,
calculated at $r$=100 AU.}
\tablenotetext{b}{Revised spectral type from \cite{dmr97}.}
\end{deluxetable*}

High-sensitivity spectra of PAHs allowing precise band strength 
measurements can provide information about the structure 
of disks and their physical and chemical conditions.  The 
presence or absence of PAH features may provide clues to the 
physical properties of the disks in HAeBe systems, such as 
their degree of flaring \citep[cf.][]{mee01}, and may also serve 
as an evolutionary clock that tracks grain processing, since 
there are multiple destruction paths for PAHs in disks (intense 
irradiation, compaction and incorporation in larger grains) but 
no clear formation path.  The PAHs in disks around HAeBe stars 
most likely originate in the ISM from which the stars formed.  
As a significant interstellar component, PAHs may have condensed 
onto the ice mantles of dust grains in dense clouds that were 
then incorporated into protostellar nebulae.  The PAHs emerge as 
free-flying molecules when the ice mantles of grains sublimate 
due to stellar irradiation or grain-grain collisions 
\citep{ll03}.  The origin of interstellar PAHs, however, is not 
well constrained.  Suggested sources for interstellar 
PAHs include: (1) formation around and ejection from carbon stars 
\citep{lat91}, (2) shattering of carbonaceous interstellar dust, 
or of photoprocessing of mantles on interstellar dust grains 
\citep{gre00}, by grain-grain collisions in interstellar 
shocks \citep{jon96}, or (3) {\it in-situ} formation through 
ion-molecule reactions \citep{her91}.  It is essential to 
quantify the PAH features as a diagnostic in order to 
distinguish between these possibilities.  

We have begun a spectroscopic survey in the mid-infrared 
(5.2--36~\mum) of 16 HAeBe stars from the sample defined by 
\cite{mal98} using the Infrared Spectrograph 
\citep[IRS;][]{hou04} on the {\it Spitzer Space Telescope} 
\citep{wer04}.  One of the goals of this study is to closely 
examine the PAH emission in the 5.2--14~\mum\ spectral 
region.  We present the first results of our study with 
spectra and analysis for four stars that show PAH features, 
no silicate features, and few other solid state features in 
their spectra. This has allowed a careful examination of 
the relative positions, strengths, and shapes of the features
at 6.2, 7.7--7.9, 11.3, and 12.7~\mum\ without concern for the 
effects of silicate dust grains, either amorphous or crystalline. 
While {\it ISO}/SWS and PHOT-S spectra of the four sources in
our sample show PAH emission features \citep{mee01, aa04}, the 
{\it Spitzer} IRS spectra have higher sensitivity, allowing the 
detailed analysis that we present here. 

All four of our sources show evidence of circumstellar disks 
in their SEDs \citep{mal98}.  HD 34282 evidently has a large 
Keplerian disk inferred from millimeter observations 
\citep{mer04, pie03}, and the disk around HD 141569 has 
been imaged directly in the optical and near infrared 
\citep{boc03, mou01, wei99, aug99, wei05}.  HD 135344 has 
yet to be spatially resolved \citep{mee01}, but \cite{hab05}
have spatially resolved the PAH emission at 3.3~\mum\ in
HD 169142.

\section{Observations and Analysis} 

The four HAeBe targets were observed with the IRS on {\it
Spitzer} in 2004 from January to October, as described in 
Table 1.  This paper focuses on the observations with the 
Short-Low module (SL), which covers the 5.2--14~\mum\ 
region with a spectral resolving power of R$\sim$90.  The
SL slit is 3\farcs6 (2 pixels) across and 57\arcsec\ long.


We started with the flatfielded images generated by the 
standard IRS data reduction pipeline (S11.0) at the Spitzer 
Science Center (SSC).  We removed the background emission by
differencing images with the source in the separate nod 
positions of each of the SL apertures.  When extracting spectra 
from the images, we used an extraction window which scales with 
wavelength; this method is similar to that used by the SSC 
pipeline.  We obtained a spectrophotometric calibration by 
using identically extracted spectra of the standard star 
HR 6348 (K0 III) and spectral templates of this star prepared 
using the methods described by \cite{coh03} and modified as 
described by \cite{slo05}.  The spectra in both nods and in 
both SL apertures were calibrated separately and then combined 
into one spectrum.  This method generally results in very high 
precision within a single spectrum, indicated by the error 
bars in our plots, and an absolute photometric accuracy of 
$\sim$10\% or better.

We have modified the S11.0 version of the wavelength 
calibration, as described by \cite{irs05}.  This modification
shifts the wavelength calibration in SL order 1 by 0.04~\mum\
to the blue.  We estimate that our wavelengths are accurate to 
better than 0.02~\mum.


Figure 1 presents the spectra of our four HAeBe stars (with 
the spectrum of HD 135344 divided by 10 to fit on the same 
plot with the other three).  The figure includes error bars 
which are generally smaller than the vertical widths of the 
lines used to plot the spectra.  These errors are the formal 
uncertainty in the mean of the spectra from the two nod 
positions in the SL slit (half the absolute value of the 
difference).  With the notable exception of HD 141569, which 
appears to have a single cool component peaking at 40~\mum, 
the SEDs in our sample consist of a combination of a 
warm component peaking in the near infrared ($\sim$2~\mum) 
and a cooler component peaking at longer wavelengths 
\citep{mal98}.  This double-peaked structure may indicate 
that the disks have formed significant gaps between a warm 
inner disk and a cool outer disk. The wavelength of the dip 
in intensity between the warm and cool components shifts 
slightly from star to star. Thus three of the spectra are 
flat because the wavelength region observed by SL lies more 
in the dip between these two components.  The continuum of 
HD 135344 appears blue because its dip appears towards longer
wavelengths than the other sources and is therefore dominated 
by the warm component ($T\sim1000$ K) at the short wavelength 
end of our spectrum.  

\begin{figure} 
\includegraphics[width=3.5in]{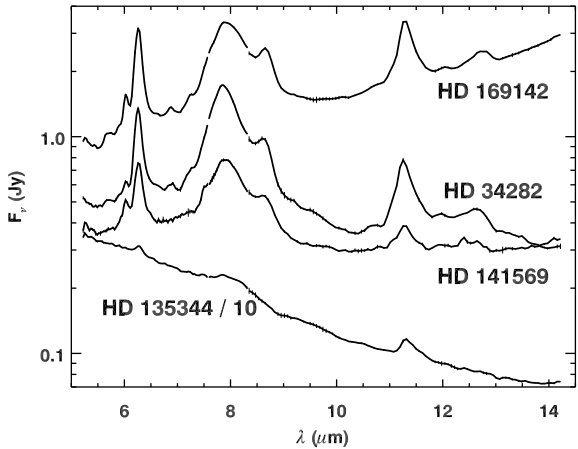}
\caption{Spectra of the sample of Herbig AeBe stars showing PAH 
emission features and no features from silicate dust grains in 
their spectra.  The spectrum of HD 135344 has been divided by 10.
These spectra do include error bars, although they are generally
smaller than the width of the lines used to plot the data.}
\end{figure}

\begin{figure} 
\includegraphics[width=3.5in]{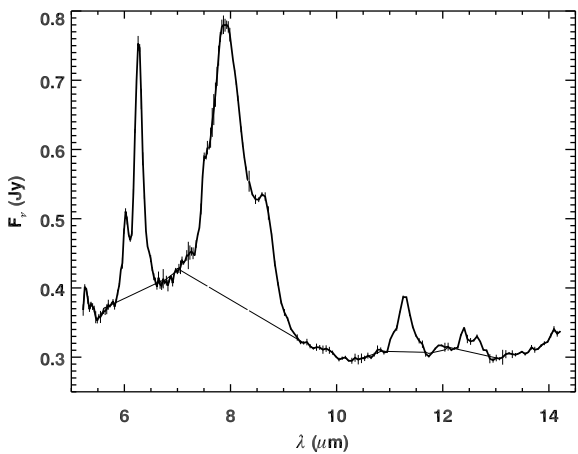}
\caption{The spectrum of HD 141569 showing the linear fitting 
method used to measure the equivalent flux in the PAH features.
Table 2 gives the wavelengths used to anchor the line segments on
either side of the PAH features.}
\end{figure}

\begin{deluxetable}{ccc}
\tablecolumns{3}
\tablewidth{0pt}
\tablenum{2}
\tablecaption{Fitting wavelengths}
\tablehead{
  \colhead{Feature (\mum)} & \colhead{$\lambda_{blue}$ (\mum)} &
  \colhead{$\lambda_{red}$ (\mum)} }
\startdata
6.2    &  5.60--5.90  &  6.71--6.78 \\
7.9    &  6.98--7.07  &  9.20--9.40\tablenotemark{a} \\
11.3   & 10.83--10.90 &  11.70--11.86 \\
12.7   & 12.10--12.30 &  12.97--13.10 \\
\enddata
\tablenotetext{a}{8.85--9.05~\mum\ for HD 135344.}
\end{deluxetable}


Figure 2 illustrates how we measure the strengths of the PAH 
features, using HD 141569 as an example.  We have used line 
segments as a rough estimate of the continuum under each 
feature.  We anchor the line segment to the spectrum on 
either side of each feature and integrate the residual 
spectrum between the two anchor points to determine the 
total flux in the feature (in W m$^{-2}$).  Table 2 lists
the wavelength intervals used to fit each feature.  This
method assumes that the PAH emission is optically thin, and 
it avoids dependencies on more sophisticated techniques to 
model the continuum.  HD 135344 is the exception; before
extracting the features using line segments, we first
fit a polynomial estimate of the continuum to the spectrum 
between the features and subtract it.

We have avoided the problem of separating the PAH features 
from other dust features like amorphous silicate emission 
at 10~\mum\ or crystalline silicate emission at 11.2~\mum\ 
by choosing our sample carefully.  The primary uncertainty 
is how much of the emission {\it below} the line segments 
belongs to the underlying PAH emission plateaux and how much 
belongs to the feature we are measuring.  If we are 
truncating flux from a PAH feature, we are doing so 
consistently to all of the stars in the sample, and our 
relative measurements should remain valid.

The resulting profiles for the PAH features at 6.2, 
7.7--7.9, 11.3, and 12.7~\mum\ appear in Figures 3 and 4.  
Table 3 presents the total fluxes and central wavelengths 
for each PAH feature in each of our sources.  The nominal
7.7--7.9~\mum\ feature is shifted to 7.9~\mum\ (or further)
in our sample, so for the remainder of the paper, we will
refer to this feature as the 7.9~\mum\ feature.  The
reader should note that the flux labelled $F_{7.9}$ in 
Table 3 and plotted in following figures includes the 
emission of the 8.6~\mum\ feature as well.  It could be 
more properly described as the total PAH flux from 
7 to 9~\mum.  We estimated the central wavelength of each 
feature by finding the wavelength at which half the emission 
lies to the blue and the red and by fitting Gaussians.  
The two methods produce similar results, but we adopt the 
Gaussian fits because they are less susceptible to noise 
and blended features.

\begin{figure} 
\includegraphics[width=3.5in]{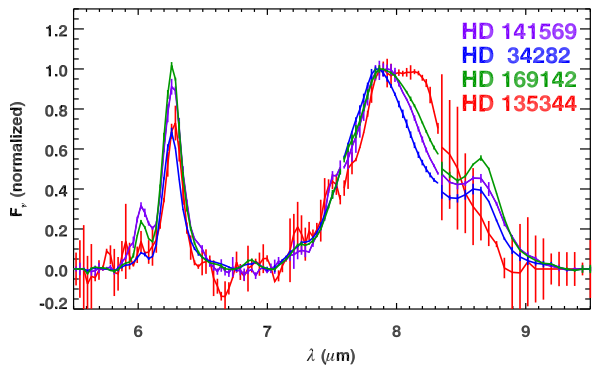}
\caption{The 6.2~\mum\ PAH emission feature and the 7--9~\mum\ 
PAH emission complex (with error bars), all normalized to the 
peak emission in the vicinity of 8~\mum.  Line segments have 
been fit under the two emission regions and subtracted.  In all 
four spectra, the ``6.2~\mum'' feature has shifted to 6.3~\mum, 
and the peak emission around 8~\mum\ occurs at (or beyond) 
7.9~\mum\ and not 7.7~\mum\ as often seen in other objects.  
The spectrum of HD 135344 shows an emission plateau from 8.0 to
8.2~\mum\ and is missing the 8.6~\mum\ feature.}
\end{figure}

\begin{figure} 
\includegraphics[width=3.5in]{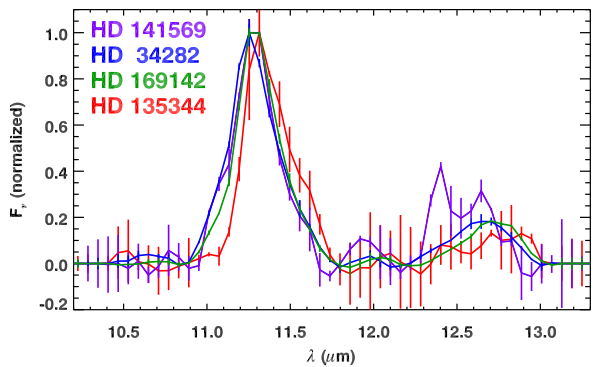}
\caption{The PAH emission features at 11.3 and 12.7~\mum\ (with 
error bars), after continuum subtraction.  Separate line segments 
have been fit under the features and removed.  All spectra are 
normalized at 11.3~\mum.  The most ionized spectrum, HD 141569, 
shows an emission feature at 12.4~\mum.}
\end{figure}

\begin{deluxetable*}{lllllllll}
\tablecolumns{9}
\tablewidth{0pt}
\tablenum{3}
\tabletypesize{\small}
\tablecaption{Feature positions and strengths}
\tablehead{ 
  \colhead{ } & \multicolumn{2}{c}{6.2~\mum\ feature}
              & \multicolumn{2}{c}{7.9~\mum\ feature} 
	      & \multicolumn{2}{c}{11.3~\mum\ feature} 
	      & \multicolumn{2}{c}{12.7~\mum\ feature}\\
  \colhead{ } & \colhead{$\lambda$} & \colhead{Flux} 
              & \colhead{$\lambda$} & \colhead{Flux} 
              & \colhead{$\lambda$} & \colhead{Flux} 
              & \colhead{$\lambda$} & \colhead{Flux} \\
  \colhead{Target} 
  & \colhead{(\mum)} & \colhead{(10$^{-15}$ W m$^{-2}$)}
  & \colhead{(\mum)} & \colhead{(10$^{-15}$ W m$^{-2}$)}
  & \colhead{(\mum)} & \colhead{(10$^{-15}$ W m$^{-2}$)}
  & \colhead{(\mum)} & \colhead{(10$^{-15}$ W m$^{-2}$)}
}
\startdata
HD  34282 &  6.26 & 12.63 \er 0.17 &  7.92 & 47.09 \er 0.58 &
            11.27 &  3.11 \er 0.14 & 12.64 &  0.54 \er 0.06 \\
HD 141569 &  6.26 &  6.84 \er 0.15 &  7.98 & 16.65 \er 0.35 &
            11.28 &  0.62 \er 0.10 & 12.52 &  0.18 \er 0.02 \\
HD 169142 &  6.26 & 32.71 \er 0.54 &  8.01 & 88.06 \er 0.93 & 
            11.30 & 11.86 \er 0.58 & 12.71 &  1.95 \er 0.17 \\
HD 135344 &  6.28 &  3.47 \er 0.51 &  8.02 & 12.49 \er 1.30 &  
            11.36 &  1.38 \er 0.21 & 12.75 &  0.18 \er 0.06 \\
\enddata
\end{deluxetable*}

Since ultraviolet and visible photons from the stellar 
photosphere excite the PAH emission \citep{ld02}, we have 
estimated integrated fluxes for each of our program stars, 
relative to the interstellar radiation field (ISRF) in the 
Solar neighborhood \citep{mm83}.  These estimates are 
intended only as a guide in our discussions of PAH 
excitation.  \cite{ll03} calculated the integrated flux 
($\lambda$=0.09--1.00~\mum) for HD 141569 as a function of 
distance, $r$, from the central star using spectral type 
B9.5 V and a Kurucz photospheric model.  Table 1 presents 
the integrated fluxes of our stars for $r$=100 AU calculated 
relative to their value for HD 141569.  Following the 
nomenclature of \cite{ll03}, $U(r)=1$ corresponds to the 
ISRF in the solar neighborhood.  We note that these are 
{\em not} in units of Habings or G$_0$, the ratio of the UV 
energy density to the estimate of Habing \citep{ha68}, since 
we include the visible spectrum in addition to UV.  Because
the PAHs are emitting either from the surface of the disk or
from the surrounding envelope, we assume no absorption 
between the photosphere and the PAH molecules.

\section{Discussion} 

\subsection{General spectral characteristics} 

The 7.7~\mum\ PAH feature consists of two main components, 
one at 7.65~\mum\ and the other at 7.85~\mum\ \citep{coh89, 
bre89, pee02}.  The 7.65~\mum\ feature tends to dominate the 
PAH spectra of reflection nebulae and H II regions where the 
PAHs seem to have been heavily processed.  The 7.85~\mum\ 
feature dominates in many planetary nebulae and objects 
evolving away from the asymptotic giant branch (AGB) where 
the PAHs seem to be relatively fresh and unprocessed.  
\cite{pee02} studied 57 spectra of PAH sources obtained with 
the SWS, and they found that in spectra where the 7.85~\mum\ 
component dominated, the 6.2~\mum\ PAH feature was shifted to 
6.3~\mum.  They used these two characteristics as the basis 
for what they called Class B PAH spectra.  

Of the 57 spectra studied by \cite{pee02}, 42 belong to 
Class A, which shows the PAH emission features at 6.2 and 
7.7~\mum.  Class B includes 12 spectra, and Class C, where 
the peak emission in the 7--9~\mum\ range is shifted to the
red of 8.0~\mum, accounts for 2 sources.  One source in
their sample had characteristics of both Class A and B.

\begin{figure} 
\includegraphics[width=3.5in]{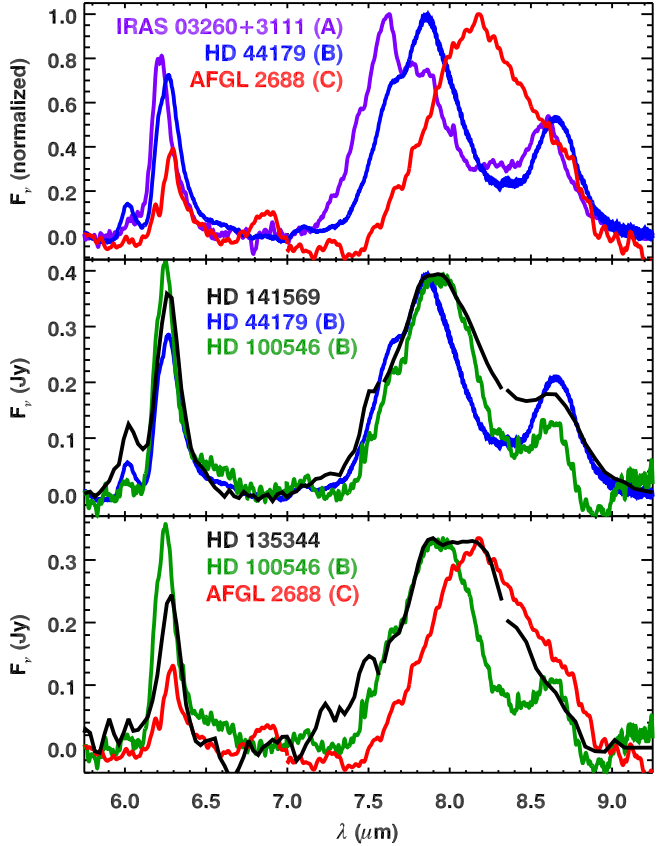}
\caption{A comparison of spectra from the SWS on \iso\ with two 
spectra from our sample.  Top:  Prototypical SWS spectra of the 
three classes defined by \cite{pee02}.  IRAS 03260+3111 is a 
young stellar object, while HD 44179 (the Red Rectangle) and 
AFGL 2688 (the Cygnus Egg) are post-AGB objects.  Middle:  
HD 141569 compared to two Class B objects.  HD 141569 resembles 
the HAeBe star more than HD 44179.  Bottom:  HD 135344 compared 
to HD 100546 and the Class C spectrum from AFGL 2688, showing 
that the spectrum resembles a blend between the two.}
\end{figure}

Figure 5 compares the spectra of HD 141569 and HD 135344
in our sample to some of the sources considered by 
\cite{pee02}.  The top panel illustrates PAH spectra of 
Classes A, B, and C as defined by \cite{pee02}, using as 
prototypes IRAS 03260+3111 (a young stellar object in the 
NGC 1333 molecular cloud core), HD 44719, (the Red 
Rectangle), and AFGL 2688 (the Cygnus Egg), respectively.
The comparison spectra were obtained from the SWS Atlas 
\citep{slo03}.  All of the IRS spectra and the SWS spectra
of IRAS 03260 and HD 44179 have had the continuum removed
using line segments as illustrated in Figure 2.  The SWS
spectra of AFGL 2688 and HD 100546 have steep red continua;
to isolate the PAH spectrum we first fit a polynomial to 
the continuum and subtracted it.

Figure 3 and the middle panel of Figure 5 show that our
HAeBe stars have spectral features typical of Class B, with 
the 6.2~\mum\ feature shifted to at least 6.26~\mum, and 
the peak emission in the 7.6--7.9~\mum\ region falling near 
7.9~\mum.  Both the 6.2 and 7.9~\mum\ features arise from 
C$-$C stretching modes.  The 8.6~\mum\ feature, which is 
included in our measurements of the strength of the 7.9~\mum\ 
feature, probably arises from a C$-$H in-plane bending mode.  

The sample examined by \cite{pee02} includes two isolated
HAeBe stars, HD 100546 and HD 179218, both of which they 
included with the Class B PAH spectra.  The middle panel
of Figure 5 shows that in the spectrum of HD 100546, the 
7.9~\mum\ PAH feature is actually shifted to even longer
wavelengths, nearly to 8.0~\mum.  In the other HAeBe star,
this feature appears at 7.8~\mum, but in all four of the
IRS spectra, this feature lies closer to 8.0~\mum.  Thus, 
in five of the six isolated HAeBe stars studied here or by
\cite{pee02}, the 7.8--7.9~\mum\ feature indicative of 
Class B is shifted nearly to 8.0~\mum.  The HAeBe stars 
may represent a subgroup within Class B.

Other features visible in the spectra in Figure 1 in the
5--8~\mum\ range require comment as well.  The 5.2~\mum\ 
PAH emission feature appears in three of the spectra at 
the short-wavelength cut-off, but it is absent in HD 34282.  
Two of the spectra, HD 34282 and HD 169142, show spectral 
structure in the vicinity of 7~\mum\ which is due to
emission features at 6.85 and 7.25~\mum.

\cite{chi00} made the first detection of absorption features
at 6.85 and 7.25~\mum, which they attributed to C$-$H
stretching modes in aliphatic hydrocarbons such as methyl
or methylene groups.  This discovery was made in the SWS
spectrum of Sgr A*.  \cite{bou01} detected the 6.85~\mum\
band in emission in the HAeBe star HD 163296.  \cite{spo04}
recently reported the 6.85 and 7.25~\mum\ bands in absorption
in the spectrum of the ultraluminous infrared galaxy IRAS
F00183-7111, and they showed that laboratory data of
hydrogenated amorphous carbon \citep[HACs;][]{fur99}
provide a good match.  We note that the 3.4~\mum\ interstellar
absorption band, identified as an aliphatic C$-$H stretching
mode \citep{san91}, can also appear in emission 
\citep{gvw90,geb92}.  It is likely, therefore, that we have 
observed these aliphatic HAC features in emission in 
HD 34282 and HD 169142.

\subsection{HD 135344} 

The spectrum of HD 135344 contains a much stronger 
contribution from a warm  ($T\sim1000$ K) underlying the PAHs, 
making the shape and position of the weaker PAH features 
much more susceptible to any systematic problems with the 
extraction and calibration of the spectrum, which explains 
the larger error bars in its spectrum in Figure 3.  
Nonetheless, some differences from the other spectra in the 
sample are still apparent. 

In the 7--9~\mum\ region, HD 135344 shows a plateau extending
from 8.0 to 8.2~\mum, most likely a blend of peaks at these
two wavelengths.  The 8.0~\mum\ peak is similar to that seen 
in four of the other HAeBe stars, but the 8.2~\mum\ peak more
closely resembles Class C PAH spectra \citep{pee02}.  Both of 
the objects in this class, AFGL 2688 and IRAS 13416-6243, are 
post-AGB objects, and neither show an 8.6~\mum\ feature.  
HD 135344 has more emission shortward of 8.0~\mum\ than these 
two objects, but the missing 8.6~\mum\ feature is striking.  

The spectrum of HD 135344 differs from the other HAeBe stars 
in other ways.  The 11.3~\mum\ feature is shifted by at least
0.06~\mum\ to the red; this is too large to arise from the
random pointing errors which occasionally place a star off the
central axis of the IRS slit.  HD 135344 also has a relatively
weak contribution 6.0~\mum\ (compared to 6.3~\mum).

\cite{bei96} attributed the 6.0~\mum\ satellite of the 
6.2~\mum\ feature to a weak C$-$C mode, but we note that in 
the objects studied by \cite{pee02} the strength of the 
6.0~\mum\ feature shows no correlation with that of the 
6.2~\mum\ C$-$C mode.  A correlation between the strength of 
the two bands also seems absent in our sample.  \cite{pee02} 
speculated that the 6.0~\mum\ feature arises from C$-$O 
stretching modes in oxygenated PAH species, but these modes 
generally peak at somewhat shorter wavelengths \citep{shu90}.  
The carrier of the 6.0~\mum\ feature remains uncertain.

\cite{oud92} have revised the spectral classification of 
HD 135344 from A0 V \citep{mss82} to F4 Ve, and \cite{dmr97}
have confirmed this new spectral type.\footnote{\cite{zuc95}
and \cite{cw95} find even later spectral types, F6 V and F8 V,
respectively.  \cite{cw95} point out that HD 135344 is a
visual binary with the two components, SAO 206462 and 206463,  
separated by 20\arcsec.  SAO 206463 is the brighter source
and is probably the origin of the A0 V classification, but
the coordinates of HD 135344 match those of the fainter 
SAO 206462, which is the target under consideration here.}
It is interesting that this source, which is significantly 
cooler than the other three sources in the sample, is the
one which shows the most unusual PAH emission in the 8~\mum\
region.

\subsection{Dependence of PAH emission on ionization fraction} 

The relative strengths of PAH features can be used to 
identify the degree of ionization of the PAHs.  They may 
therefore provide clues to the incident radiation field,
electron densisty, and gas temperature.  The most obvious 
difference between neutral and ionized PAHs is the relative 
strength of the C$-$C modes in the 7.7--7.9~\mum\ region to 
the out-of-plane C$-$H bending modes at 11.3~\mum\ and 
longer wavelengths \citep[e.g.][]{ahs99}.  Ionized PAHs emit 
more strongly in the features at 6--9~\mum\ than in the 
10--13~\mum\ region, while in neutral PAHs, the 10--13~\mum\ 
region dominates.  The models of \cite{dl01} show that the
ratio $F_{7.9}/F_{11.3}$ increases with the size of the PAH
molecule since larger PAHs generally have a higher C:H ratio
\citep[e.g. see][]{ld01}.  However, a shift from neutrals to 
ions increases the ratio much more significantly, by a factor 
of $\sim$10.

We can thus use the ratio of the strengths of the 7.9 and 
11.3~\mum\ features as an estimate for the relative degree 
of ionization in each spectrum.  This ratio is the most 
obvious difference among the spectra in our sample, varying 
from 7.4 in HD 169142 to 25.2 in HD 141569, making these 
the least and most ionized PAH spectra, respectively.  The 
ratios for the intermediate stars are 9.1 for HD 135344 and 
15.2 for HD 34282.

Recent optical spectroscopy by \cite{dmr97} confirms the
previous revision of the spectral classification of HD 135344,
as discussed above, and also revises the classification of
HD 169142 from B9 V \citep{mss82} to A5 Ve.  These new
classifications correlate with the observed ionization
fractions as measured by the $F_{7.9}/F_{11.3}$ ratio, with 
the two A0 stars having the more ionized spectra, and the
later-type stars having the less ionized spectra.

In HD 141569, according to detailed models \citep{ll03}, the
charging of PAHs is dominated by collisions with electrons 
and not by emitting photoelectrons.   The electron 
recombination rate is three orders of magnitude larger than 
the photoelectron emission rate.  For this case, then, the 
PAHs are negatively charged.  Our analysis is not sensitive 
to the polarity of the charge of the PAHs, only the ionization
fraction.

\subsection{The 7.9~\mum\ feature} 

There is no clear explanation for the varying strengths of 
the components of the 7--9~\mum\ PAH complex among observed 
sources.  \cite{bt05} investigate two possible causes in 
their study of the spatial distribution of the components in 
extended reflection nebulae.  The peak of the complex might 
shift from 7.6 to 7.9~\mum\ as the fraction of PAH cations 
increases or as the fraction of fresh and unprocessed PAHs 
increases.  Our sample shows a range of ionization fractions, 
but in three of the four spectra, the shape of the 7--9~\mum\ 
complex looks similar.  The ratio of the 7.9~\mum\ and 
11.3~\mum\ features varies by a factor of 3.7 in these three 
sources, and yet the relative contributions of the 7.6 and 
7.9~\mum\ components seems fixed, which would suggest that
the components are not sensitive to ionization, at least
over the ionization range we have examined.  

The size of the PAH molecules may also influence the 
relative strengths of the features at 7.6 and 7.9~\mum, but 
there is some disagreement over how the position of the
strongest feature in this wavelength range will depend
on PAH size.  \cite{bjp01} suggest that very small grains 
produce the longer-wavelength feature, but \cite{hud99b} 
report that their laboratory measurements show that the 
7.6~\mum\ feature shifts to longer wavelength for {\it 
larger} PAHs.  We can only say that our data are least
consistent with the dependence of the 7--9~\mum\ PAH
complex on ionization fraction. 

\subsection{Possible correlations with PAH size and structure} 

The models of \cite{dl01} show a dependence of the ratio 
$F_{6.2}/F_{7.9}$ on the size of the PAH molecule, but it 
is weak, with the flux ratio doubling only when the number 
of C atoms has been reduced by a factor of $\sim$10.  The 
top panel of Figure 6 shows that the ratio $F_{6.2}/F_{7.9}$ 
varies by a factor of two in our sample, but not in any 
systematic way with the ionization fraction as indicated 
by $F_{7.9}/F_{11.3}$.

\begin{figure} 
\includegraphics[width=3.5in]{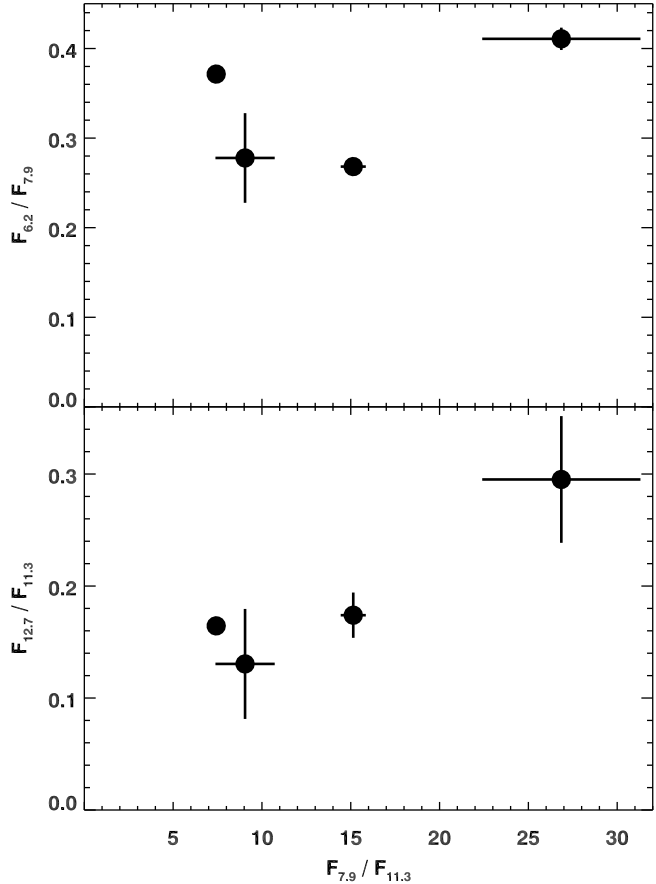}
\caption{Flux ratios from the different PAH features plotted vs. 
the ratio of the 7.9 and 11.3~\mum\ PAH features.  The ratio 
$F_{7.9}/F_{11.3}$ traces the ionization fraction of the PAHs.  
From left to right (least to most ionized), the symbols are for 
HD 169142 (A5), HD 135344 (F4), HD 34282 (A0), and HD 141569 
(A0).  In the top panel, the relative strengths of the 6.2 and 
7.9~\mum\ features vary by a factor of $\sim$2, but they do not 
vary systematically with ionization fraction.  The bottom panel 
shows that the ratio $F_{12.7}/F_{11.3}$ increases with 
ionization fraction, suggesting that the size of the PAHs 
decreases with ionization fraction.}
\end{figure}

To better discriminate the size of the PAH molecules, 
\cite{hon01} use the ratio $F_{12.7}/F_{11.3}$.  The 
11.3~\mum\ feature arises from a C$-$H stretching mode
when an aromatic ring contains only one H atom, such as
one might find along the edge of a large PAH molecule.
The 12.7~\mum\ feature arises from the C$-$H trio 
stretching mode, i.e. from aromatic rings with three H
atoms, which will be found on the corners and ends of PAH
molecules.  As one increases the size of a PAH molecule,
the strength of the emission at 11.3~\mum\ will increase
with respect to the emission at 12.7~\mum.

The bottom panel of Figure 6 shows that the strength 
of the 12.7~\mum\ feature generally increases with the 
ionization fraction.  Comparing this panel with the top 
panel reveals that the use of the ratios $F_{6.2}/F_{7.9}$ and 
$F_{12.7}/F_{11.3}$ as size indicators gives contradictory 
results.

The $F_{12.7}/F_{11.3}$ ratio may be influenced not just
by the size of the PAHs, but by their geometry.  If a 
compact PAH is rearranged into a structure with the same 
number of carbon atoms but with more uneven edges, the
strength of the duo and trio modes will increase compared 
to the solo mode.  Thus, it is possible that the the 
enhancement of the 12.7~\mum\ feature could also arise 
from more irregularly shaped and less compact PAHs.  

Measurements at the Astrochemistry Laboratory at NASA Ames 
Research Center have revealed no dependency of the 
$F_{6.2}/F_{7.9}$ ratio on the size of the PAHs (Hudgins 
2004, private communication), but the Astrochemistry Laboratory 
has not explored a full order of magnitude in PAH size.  
While larger PAHs, such as those modeled by \cite{dl01},
do show a dependence, typical PAH sizes are on the order of
$\sim$100 carbon atoms \citep[e.g.][]{ld01}, and the range in 
PAH sizes may be insufficient to dominate other causes for 
variation in the strength of the 6.2 and 7.9~\mum\ bands.  The 
relative strengths of the out-of-plane C$-$H bending modes, on 
the other hand, are much more sensitive to variations in PAH 
size, especially for the sizes of the molecules likely to 
be encountered in typical astrophysical environments.  

The spectra of our HAeBe stars show other emission features
in the 11--13~\mum\ region in addition to the 11.3 and 
12.7~\mum\ features.  In Figures 1 and 4, the duet out-of-plane 
C$-$H bending mode at 11.9~\mum\ is visible in all of the 
spectra except HD 135344.  These figures also show
that what we are measuring as the 12.7~\mum\ feature 
shows some structure.  In the spectrum of HD 141569, the 
secondary emission peak at 12.4~\mum\ actually dominates 
the 12.7~\mum\ feature.  The Humphreys-$\alpha$ line should
be at 12.365~\mum, but there is no corresponding Pfund-$\alpha$ 
line at 7.46~\mum.   The H$_2$ 0--0 S(2) is expected 
at 12.28~\mum, but this is too far to the blue and it is 
not accompanied by the 0--0 S(3) line at 9.67~\mum.  The 
most likely carrier of the 12.4~\mum\ feature is another 
trio out-of-plane bending mode, as this feature can vary 
from 12.1 to 13.5~\mum, depending on the PAH molecule 
\citep[e.g.][]{ha99}.

\section{Conclusion} 

We have presented mid-infrared spectroscopy from {\it Spitzer} 
of four Herbig AeBe stars whose spectra show strong PAH 
emission but no silicate dust features.  The absence of 
silicate emission has enabled a detailed analysis 
independent of specific disk models of the relative 
strengths and positions of PAH emission features in the 
5--14~\mum\ region.  

Our sample shows spectral characteristics generally 
consistent with the Class B PAH emission sources identified 
by \cite{pee02}, viz. a 6.2~\mum\ feature shifted to 
6.3~\mum\ and 7--9~\mum\ emission dominated by the 
component at 7.9~\mum.  However, in all of our sources, the 
7.9~\mum\ feature is shifted closer to 8.0~\mum, as in the 
spectrum of HD 100546, a HAeBe star observed by the SWS on 
\iso.

The spectra show a range in ionization fraction as
revealed by the $F_{7.9}/F_{11.3}$ ratio, with HD 141569 
showing the highest degree of ionization and HD 169142 the 
lowest.  The ionization fraction is lower for 
later spectral classes.  
The strength of the 12.7~\mum\ feature increases with 
respect to the 11.3~\mum\ feature as the ionization 
fraction increases.
The strength of the 6.2~\mum\ feature relative to the
7.9~\mum\ feature varies by a factor of nearly two in
our sample, but in no systematic way.

The spectrum of HD 135344, the coolest star in our sample, 
resembles the Class C spectra discussed by \cite{pee02} 
in some ways, with more emission to the red of 7.9~\mum\ 
than in the other sources and a missing 8.6~\mum\ feature.

Thus, our sample of PAH spectra from HAeBe stars show 
some characteristics which reinforce results from the 
literature, but also raise new questions.  Why do 
HAeBe stars consistently produce the Class B PAH spectra 
seen in less than 25\% of the SWS sample?  What causes 
the shift of the 7.9~\mum\ feature to $\sim$8.0~\mum?  
What causes the range of 6.2~\mum\ feature strengths 
compared to the 7.9~\mum\ feature?  What is the carrier 
of the 6.0~\mum\ feature?  Answering these questions will 
help us understand the nature of the HAeBe systems from 
which the PAH emission arises.  

Quantitative measures of the size and ionization fraction 
of the PAHs would be valuable in determining the degree 
of grain processing in the disks and the fraction of 
short-wavelength stellar photons that reach the outer 
disks.  With the small size of the current sample, it is 
not possible to determine the origin of the correlations, 
although we expect that the full IRS sample of HAeBe stars
will bring us closer to this goal.

\acknowledgments

J.~D. Bregman, D.~M. Hudgins, and L.~J. Allamandola provided 
useful comments as we prepared this manuscript.  
This work is based on observations made with the Spitzer 
Space Telescope, which is operated by the Jet Propulsion 
Laboratory, California Institute of Technology under NASA 
contract 1407.  Support for this work was provided by 
NASA through contract number 1257184 issued by JPL/Caltech.
This research has made use of the SIMBAD database operated 
at the Centre de Donn\'{e}es astronomiques de Strasbourg.


\begin{thebibliography}{}

\bibitem[Acke \& van den Ancker(2004)]{aa04} Acke, B. \& van den 
  Ancker, M.~E.  2004, \aap, 426, 151
\bibitem[Allamandola et al.(1999)]{ahs99} Allamandola, L.~J., Hudgins,
  D.~M., \& Sandford, S.~A. 1999, \apj, 511, L115
\bibitem[Allamandola et al.(1989)]{atb89} Allamandola, L.~J., Tielens,
  A.~G.~G.~M., \& Barker, J.~R. 1989, \apjs, 71, 733
\bibitem[Augereau et al.(1999)]{aug99} Augereau, J.~C., Lagrange, 
  A.~M., Mouillet, \& Menard, F. 1999a, \aap, 350, L51
\bibitem[Beintima et al.(1996)]{bei96} Beintima et al. 1996, \aap, 
  315, L369
\bibitem[Boccatelli et al.(2003)]{boc03} Boccatelli, A., Augereau, 
  J.~C., Marchis, F., \& Hahn, J. 2003, \apj, 585, 494
\bibitem[Boissel et al.(2001)]{bjp01} Boissel, P., Joblin, C., \& 
  Pernot, P., 2001, \aap, 373, L5
\bibitem[Bouwman et al.(2001)]{bou01} Bouwman, J., Meeus, G., de Koter, A.,
  Hony, S., Dominik, C., \& Waters, L.~B.~F.~M.
\bibitem[Bregman(1989)]{bre89} Bregman, J.~D. 1989, in
  Interstellar Dust, Proc. IAU Symp 135, ed. L.~J. Allamandola \& 
  A.~G.~G.~M. Tielens, 109
\bibitem[Bregman \& Temi(2005)]{bt05} Bregman, J.~D. \& Temi, P. 
  2005, \apj, in press
\bibitem[Chiar et al.(2000)]{chi00} Chiar, J.~E., Tielens, 
  A.~G.~G.~M., Whittet, D.~C.~B., Schutte, W.~A., Boogert, A.~C.~A.,
  Lutz, D., van Dishoeck, E.~F., \& Bernstein, M.~P. 2000, \apj,
  537, 749
\bibitem[Cohen et al.(2003)]{coh03} Cohen, M., Megeath, T.~G., 
  Hammersley, P.~L., Martin-Luis, F., \& Stauffer, J. 2003, \aj, 125, 
  2645
\bibitem[Cohen et al.(1989)]{coh89} Cohen, M., Tielens, A.~G.~G.~M., 
  Bregman, J., Witteborn, F.~C., Rank, D.~M., Allamandola, L.~J., 
  Wooden, D., Jourdain de Muizon, M. 1989, \apj, 341, 246
\bibitem[Coulson \& Walther(1995)]{cw95} Coulson, I.~M. \& Walther,
  D.~M. 1995, \mnras, 274, 977
\bibitem[Draine \& Li(2001)]{dl01} Draine, B.~T. \& Li, A. 2001, \apj, 
  551, 807
\bibitem[Dunkin et al.(1997)]{dmr97} Dunkin, S.~K., Barlow, M.~J., \&
  Ryan, S.~G. 1997, \mnras, 286, 604
\bibitem[Furton et al.(1999)]{fur99} Furton, D.~G., Laiho, J.~W.,
  \& Witt, A.~N. 1999, \apj, 526, 752
\bibitem[Geballe \& van der Ween (1990)]{gvw90} Geballe, T.~R., \&
  van der Veen, W.~E.~C.~J. 1990, \aap, 235, L9
\bibitem[Geballe et al.(1992)]{geb92} Geballe, T.~R., Tielens,
  A.~G.~G.~M., Kwok, S., \& Hrivnak, B.~J. 1992, \apj, 287, L89
\bibitem[Greenberg et al.(2000)]{gre00} Greenberg, J.M., et al. 2000,
  \apj, 531, 71
\bibitem[Habart et al.(2004)]{hab04} Habart, E., Natta, A., \& Krugel, 
  E.  2004, \aap, 427, 179
\bibitem[Habart et al.(2005)]{hab05} Habart, E.,  Natta, A., Testi, L., 
  Carbillet, M. 2005, \aap, in press
\bibitem[Habing (1968)]{ha68} Habing, H.~J. 1968, Bull. Astron. Inst. 
  Netherlands, 19, 421
\bibitem[Herbig(1960)]{her60} Herbig, G.~H. 1960, \apjs, 4, 337
\bibitem[Herbst(1991)]{her91} Herbst, E. 1991, \apj, 366, 133
\bibitem[Hony et al.(2001)]{hon01} Hony, S., Van Kerckhoven, C., 
  Peeters, E., Tielens, A.~G.~G.~M., \& Allamandola, L.~J. 2001, \aap, 
  370, 1030
\bibitem[Houk(1982)]{mss82} Houk, N. 1982, Michigan Catalogue of
  Two-Dimensional Spectral Types for the HD Stars, Vol. 3 (Ann Arbor:
  Univ. of Michigan)
\bibitem[Hudgins \& Allamandola(1995a)]{ha95a} Hudgins, D.~M. \&
  Allamandola, L.~J. 1995a, J. Phys. Chem., 99, 3033
\bibitem[Hudgins \& Allamandola(1995b)]{ha95b} Hudgins, D.~M. \&
  Allamandola, L.~J. 1995b, J. Phys. Chem., 99, 8978
\bibitem[Hudgins \& Allamandola(1997)]{ha97} Hudgins, D.~M. \&
  Allamandola, L.~J. 1997, J. Phys. Chem., 101, 3472
\bibitem[Hudgins \& Allamandola(1999)]{ha99} Hudgins, D.~M. \&
  Allamandola, L.~J. 1999, \apj, 516, L41
\bibitem[Hudgins et al.(1994)]{hud94} Hudgins, D.~M., Sandford, S.~A.,
  \& Allamandola, L.~J. 1994, J. Phys. Chem., 98, 4243
\bibitem[Hudgins et al.(1999a)]{hud99a} Hudgins, D.~M.,\& Allamandola, 
  L.~J.  1999a, \apj, 516, L41
\bibitem[Hudgins et al.(1999b)]{hud99b} Hudgins, D.~M.,\& Allamandola, 
  L.~J.  1999a, \apj, 513, L69
\bibitem[Houck et al.(2004)]{hou04} Houck, J.~R., et al. 2004, \apjs, 
  154, 18
\bibitem[Jones et al.(1996)]{jon96} Jones, A.~P., Tielens, A.~G.~G.~M.,
  \& Hollenbach, D.~J. 1996, \apj, 469, 740
\bibitem[Latter(1991)]{lat91} Latter, W.~B. 1991, \apj, 377, 187
\bibitem[Li \& Draine(2001)]{ld01} Li, A. \& Draine, B.~T. 2001, \apj, 554, 778
\bibitem[Li \& Draine(2002)]{ld02} Li, A. \& Draine, B.~T. 2002, \apj, 572, 762
\bibitem [Li \& Lunine(2003)]{ll03} Li, A. \& Lunine, J.~I. 2003, \apj, 594, 
  987
\bibitem[Malfait et al.(1998)]{mal98} Malfait, K., Bogaert, E, \& 
  Waelkens, C. 1998, \aap, 331, 211
\bibitem[Mathis,  et al.(1983)]{mm83} Mathis, J.~S., Mezger, P.~G., \& 
  Panagia, N. 1983, \aa, 128, 212
\bibitem[Meeus et al.(2001)]{mee01} Meeus, G., Waters, L.~B.~F.~M., 
  Bouwman, J., van den Ancker, M.~E., Waelkens, C., \& Malfait, K.
  2001, \aap, 365, 476
\bibitem[Merin et al.(2004)]{mer04} Merin, B., et al. 2004, \aap, 419, 301 
\bibitem[Mouillet et al.(2001)]{mou01}Mouillet, D., Lagrange, A.~M., 
  Augereau, J.~C., \& Menard, F. 2001, \aap, 372, L61
\bibitem[Oudmaijer et al.(1992)]{oud92} Oudmaijer, R.~D., van der 
  Veen, W.~E.~C.~J., Waters, L.~B.~F.~M., Trams, N.~R., Waelkens, C., 
  Engelsman, E. 1992, \aaps, 96, 625
\bibitem[Peeters et al.(2002)]{pee02} Peeters, E., Hony, S., Van 
  Kerckhoven, C., Tielens, A.~G.~G.~M., Allamandola, L.~J.; Hudgins, 
  D.~M., \& Bauschlicher, C.~W. 2002 \aap, 390, 1089
\bibitem[Pietu et al.(2003)]{pie03} Pietu, V., Dutrey, A., \& Kahane, 
  C.  2003, \aap, 398, 565
\bibitem[Sandford et al.(1991)]{san91} Sandford, S.~A., Allamandola,
  L.~J., Tielens, A.~G.~G.~M., Sellgren, K., Tapia, M., \&
  Pendleton, Y. 1991, \apj, 371, 607
\bibitem[Shu et al.(1987)]{shu87} Shu, F.~H., Adams, F.~C., \& Lizano, 
  S. 1987, ARA\&A, 25, 23
\bibitem[Schutte et al.(1990)]{shu90} Schutte, W.~A., Tielens, A.~G.~G.~M., 
  Allamandola, L.~J., Wooden, D.~H., Cohen, M. 1990, \apj, 360, 
  577
\bibitem[Sloan et al.(1999)]{slo99} Sloan, G.~C., Hayward, T.~L., 
  Allamandola, L.~J., Bregman, J.~D., DeVito, B., \& Hudgins, D.~M., 
  1999, \apj, 513, L65
\bibitem[Sloan et al.(2003)]{slo03} Sloan, G.~C., Kraemer, K.~E., \& 
  Price, S.~D.  2003, \apjs, 147, 379
\bibitem[Sloan et al.(2005b)]{slo05} Sloan, G.~C., et al. 2005, in prep.
\bibitem[Sloan et al.(2005a)]{irs05} Sloan, G.~C., Spoon, H.~W.~W.,
  \& Bernard-Salas, J., IRS Technical Report 05002, Low-resolution 
  wavelength calibration of the IRS (Ithaca, NY:  Cornell, available 
  at http://isc.astro.cornell.edu/tech/tr/)
\bibitem[Spoon et al.(2004)]{spo04} Spoon, H.~W.~W., et al. 2004,
  \apjs, 154, 184
\bibitem[Strom et al.(1972)]{str72} Strom, S.~E, Strom, K.~M., Yost, 
  J., Carrasco, L., Grasdalen, G. 1972, \apj, 173, 353
\bibitem[The et al.(1994)]{the94} The, P.~S., de Winter, D., \& Perez, 
  M.~R. 1994, \aaps, 104, 315
\bibitem[van den Ancker et al.(2001)]{vda01} van den Ancker, M.~E., 
  Meeus, G., Cami, J., Waters, L.~B.~F.~M., \& Waelkens, C., 2001, 
  \aap, 369, L17
\bibitem[van den Ancker(2000)]{vda00} van den Ancker, M.E. 2000, in
  Disks, Planetesimals, \& Planets, ASP Conf. Series 219, ed. F.
  Garz\'{o}n, C. Eiroa, D. de Winter, \& T.~J. Mahoney, 242
\bibitem[Waters \& Waelkens(1998)]{ww98} Waters, L.~B.~F.~M \& 
  Waelkens, C. 1998, ARA\&A, 36, 233
\bibitem[Weinberger et al.(1999)]{wei99} Weinberger, A.~J., et al.
  1999, \apj, 525, L53
\bibitem[Weinberger et al.(2005)]{wei05} Weinberger, A.~J., et al. 
  2005, \apj, submitted
\bibitem[Werner et al.(2004)]{wer04} Werner, M.~W., et al. 2004, 
  \apjs, 154, 1
\bibitem[Zuckerman et al.(1995)]{zuc95} Zuckerman, B., Forveille,
  T., \& Kastner, J.~H. 1995, \nat, 373, 494

\end{thebibliography}
\end{document}